# INTERPOLATING FIELDS OF CARBON MONOXIDE DATA USING A HYBRID STATISTICAL-PHYSICAL MODEL

By Anders Malmberg[1], Avelino Arellano[1], David P. Edwards[1], Natasha Flyer, Doug Nychka and Christopher Wikle

*National Center for Atmospheric Research and University of Missouri*

Atmospheric Carbon Monoxide (CO) provides a window on the chemistry of the atmosphere since it is one of few chemical constituents that can be remotely sensed, and it can be used to determine budgets of other greenhouse gases such as ozone and OH radicals. Remote sensing platforms in geostationary Earth orbit will soon provide regional observations of CO at several vertical layers with high spatial and temporal resolution. However, cloudy locations cannot be observed and estimates of the complete CO concentration fields have to be estimated based on the cloud-free observations. The current state-of-the-art solution of this interpolation problem is to combine cloud-free observations with prior information, computed by a deterministic physical model, which might introduce uncertainties that do not derive from data. While sharing features with the physical model, this paper suggests a Bayesian hierarchical model to estimate the complete CO concentration fields. The paper also provides a direct comparison to state-of-the-art methods. To our knowledge, such a model and comparison have not been considered before.

**1. Introduction.** Atmospheric Carbon Monoxide (CO) is an important trace gas in the atmosphere. It is produced by both natural emissions and human activities and is formed primarily through natural atmospheric oxidation processes and incomplete combustion from burning fossil fuels and biomass. Although in the developed countries one can associate part of the CO production with wildfires and auto emissions, developing countries also generate CO from forest clearing and biofuels. Thus, CO is a global pollutant with a variety of sources. CO has a mean lifetime of about two months

Received December 2007; revised March 2008.
[1]Supported by NASA under Grant NNG06GB27G. The National Center for Atmospheric Research is sponsored by the National Science Foundation.
*Key words and phrases.* Carbon monoxide, satellite data, Bayesian hierarchical models, interpolation, data assimilation.

and thus can serve as a regional and global tracer of pollution and transport. CO is also related to the concentrations of OH radicals and ozone in the troposphere, two important species that are difficult to measure directly. For these reasons the concentrations of CO are a window on the troposphere, not only for tracing the motion of the atmosphere and the transport of surface emissions but also in determining budgets for other chemical constituents.

1.1. *Monitoring CO.* As one of the US Environmental Protection Agency (EPA) Criteria Pollutants, CO is regularly monitored at a sparse network of ground stations. In addition, there is a limited capability to map CO distributions from space using instruments such as the Measurement Of Pollution In The Troposphere (MOPITT) aboard the NASA Terra platform; see Edwards et al. (2004). This system is limited by the usual sampling problems from satellite based instruments and so it is difficult to synthesize complete 3-dimensional fields of CO on short time scales. An observing framework for atmospheric composition that is analogous to that achieved for weather forecasting has been recommended to improve monitoring of CO and other pollutants [Barrie, Borrell and Langen (2004)]. Currently, the missing component of such an integrated observing strategy is a platform in geostationary Earth orbit (GEO) [Edwards (2006)] that would be capable of multispectral observations with high spatial and temporal resolution. The statistical methodology in this paper addresses how to use this next generation of remote sensing platforms to estimate atmospheric CO.

A GEO platform is described as having a *stare* capability due to its fixed position. Although this sampling is an improvement over other possible orbits, remote sensing of CO is also dependent on cloud-free conditions. Thus, CO cannot be retrieved at pixels with cloud cover and so a technique for interpolation in space and time will be required to estimate the complete CO concentration field for all vertical layers from available satellite cloud-free observations. In this paper we present a new method based on physical and statistical principals.

1.2. *Data assimilation for atmospheric trace gases.* The current state-of-the-art interpolation method for chemical constituents such as CO is known in the geophysical literature as data assimilation (DA); see Lary (1999) and Kalnay (2003). Formally, DA is a Bayesian statistical method that combines a prior guess of the complete field with observations in an optimal fashion. In the simplest case under linear and Gaussian assumptions, DA methods are derived from the Kalman filter [see Kalman and Bucy (1961) and Shumway and Stoffer (2000)] and statisticians can recognize the updating step where current observations are combined with the current estimate of the state to be some variant of best linear unbiased prediction or manipulating conditional multivariate normal distributions. An important distinction in the geophysical context, such as assimilation of CO measurements, is the use of a chemical transport model (CTM) for the forecast step in the Kalman filter. Methods such as the ensemble Kalman filter and smoother actually take advantage of the physical model to constrain the covariance function used in the update step. Based on CTMs, DA systems have been applied to assimilate satellite retrievals of atmospheric chemical constituents, such as ozone, aerosols, and CO, in three-dimensional chemical transport models or CTMs [Lary (1999), Hanea, Velders and Heemink (2004), Lamarque et al. (2004), Eben et al. (2005), Sandu et al. (2005), Chai et al. (2006), Collins et al. (2006), Arellano et al. (2007), among others]. CTMs are complex physical models, and the final interpolation may contain features that are particular to the CTM rather than the true CO field. Typically, being a physical model, a CTM does not have free parameters that are adapted to the current dynamics of the CO process and so model biases and uncertainties can be hard to separate from the true features of the CO field. Uncertainties of the estimated concentrations are not easily interpreted. We believe that a simpler and stochastic model can replace the role of a CTM in assimilation, provided it is embedded in a statistical framework.

1.3. *Bayesian hierarchical models.* In contrast to the CTM-based assimilation, this paper suggests a statistical model that has the advantage of being easy to describe and provides more accurate uncertainty estimates. In particular, a Bayesian Hierarchical Model (BHM) is built by constructing conditional models for the observations and underlying process that determine the complete CO fields. Each conditional model is relatively simple, but when combined they are able to describe complex systems. Furthermore, the conditional models are designed to incorporate prior physical knowledge about the processes. This makes the approach innovative as it contains both physics and statistics. BHMs have been applied in a range of applications, for example, Wikle et al. (2001) and Wikle et al. (2003) give a detailed description of a two-dimensional model based on hybrid statistical-physical ideas.

In this work the statistical process model of CO concentration is a 3-dimensional spatial model plus temporal dynamics. To our knowledge, this complexity has not been considered before in a BHM. The dynamics of this process are motivated by a physical transport model, with some stochastic parameters introduced to reflect model uncertainties and the difficulty of resolving vertical motions at coarse resolution. It is noteworthy that the physical structure of the process model generates realistic spatial covariances that support the interpolation from cloud-free to cloudy pixels.

Another important aspect of this work is a direct comparison with a state-of-the-art assimilation from a large CTM and an ensemble Kalman filter. To our knowledge, this is the first time a BHM has been evaluated against a DA currently in use and so provides an interesting benchmark for comparing stochastic and deterministic models in the geosciences.

1.4. *Outline.* The next section presents the statistical model and the three hierarchical stages. Among these, the process stage is emphasized since it contains the description of the transport of a stochastic process on a 3-dimensional spatial grid. Here the transport is motivated by physical assumptions, merged with statistical ideas. Section 3 contains the application of the proposed model to a hypothetical GEO satellite air-quality monitoring scenario. Results and interpretations are presented together with a comparison with an existing CTM-based DA approach. Finally, Section 4 contains some remarks and suggestions for future applications and development.

**2. Hierarchical model for CO transport.** The Bayesian hierarchical model used in this application can be outlined in three levels [Berliner (1996)]:

$[Y|X,\Theta]$ Data stage,
$[X|\Theta]$ Process stage,
$[\Theta]$ Parameter stage,

where brackets are used to denote a probability density function, $Y$ represents the observation process, $X$ the underlying true CO mixing ratio on a 3-dimensional spatial grid, and $\Theta$ represents a vector of unknown parameters. The vertical line (to be read as "given") indicates a conditional distribution with the variables on the right being fixed. At the top level of this hierarchical model, the data stage specifies the likelihood of observing $Y = y$ given $X$ and $\Theta$. The next level, the process stage, models the spatio-temporal dynamics of $X$, which in our case is a hybrid model of physical equations and stochastic processes. The dynamical model involves some unknown parameters, $\Theta$, which are assigned prior distributions in the parameter stage.

Our goal is to determine the distribution of the unknowns, $X$ and $\Theta$, given the data, known as the posterior distribution. Bayes' theorem states that the posterior

$$[X, \Theta | Y = y] \tag{1}$$

is proportional to the product of the distributions in the three hierarchical stages. Typical for complicated models such as this application, the posterior does not have a closed form and must be approximated by a Monte Carlo sample. Sampling the posterior in equation (1) is implemented by Markov Chain Monte Carlo and applying a Gibbs sampler; see Robert and Casella (1999). Briefly, the idea behind the Gibbs sampler is to iteratively draw samples from the full conditional distributions of subsets of $X$ and $\Theta$. The form of the space-time model used in this work facilitates simple expressions for the full conditional distributions. A general overview of the Bayesian model approach and some applications is provided in Banerjee, Gelfand and Carlin (2004). Finally, moments and probabilities for the posterior distribution are approximated by the sample statistics for $X$ and $\Theta$, from values of the Markov chain once it appears to be stationary.

The remainder of this section details the three stages of the model, where the focus is on how a process with three spatial dimensions can be modeled hierarchically.

2.1. *Data model.* In this application we assume that the observations are obtained on the same regular grid on which the underlying process is sought. (This can be generalized without any major changes in our approach.) Furthermore, let the horizontal grid locations be denoted by $s_i$, for $i = 1, \ldots, N$, and let the vertical layers be denoted by $l$, for $l = 1, \ldots, L$, ordered from bottom to top. Finally, the concentrations are observed for equally spaced time points $k \in \{1, \ldots, T\}$.

For a given region, the amount of cloud cover will change over time along with the number of cloud-free pixels observed by the GEO satellite. For a given time point, $n$ out of $N$ locations are observed. The conditional model for the $n$-dimensional observation vector, at level $l$ and time $k$, $Y_k(l) = \{Y_k(s_j, l), j = 1, \ldots, n\}$, given the $N$-dimensional process vector, $X_k(l) = \{X_k(s_i, l), i = 1, \ldots, N\}$, is

$$(2) \qquad Y_k(l) = \mathsf{D}_k X_k(l) + \epsilon_k(l),$$

where $\epsilon_k(l) \in N(0, \Sigma_{\epsilon_k}(l))$ with $\Sigma_{\epsilon_k}(l) = \text{diag}(\sigma^2_{\epsilon_k}(s_i, l), i = 1, \ldots, n)$, and $\sigma_{\epsilon_k}(s_i, l)$ is a prescribed measurement noise taken to be 10% of $X_k(s_i, l)$; see Pan et al. (1998). Element $(j, i)$ in the incidence matrix $\mathsf{D}_k$ is 1 if element $j$ in $Y_k(l)$ represents an observation of element $i$ in $X_k(l)$.

2.2. *Process stage.* Atmospheric CO has a global average lifetime of about 2 months; see Cicerone (1988). It is envisaged that GEO observations of the process will be available hourly, on time-scales of several orders of magnitude shorter than the CO lifetime. Hence, the process can be assumed to be nonreactive at the time-scales considered here and so winds can transport CO a substantial distance, and any process model should account for transport. For example, Asian emissions can effect US regional air quality. In a continuous formulation, transport (also termed *advection*) of a nonreactive tracer by winds is described with an advection-diffusion equation, as follows:

$$\frac{\partial X(s,t)}{\partial t} = u(s,t) \frac{\partial X(s,t)}{\partial s_x} + v(s,t) \frac{\partial X(s,t)}{\partial s_y} + w(s,t) \frac{\partial X(s,t)}{\partial s_z}$$

(3)
$$+ \Theta(s)\biggl(\frac{\partial^2 X(s,t)}{\partial s_x^2} + \frac{\partial^2 X(s,t)}{\partial s_y^2} + \frac{\partial^2 X(s,t)}{\partial s_z^2}\biggr),$$

where $s = (s_x, s_y, s_z)$ for the time being is a coordinate in a three-dimensional Euclidean space, $u$, $v$, and $w$, are the three wind components (east-west, north-south, and vertical) of the actual wind field, and $\Theta(s)$ is a diffusion coefficient. Equation (3) forms the basis of the process stage, and is a rigorous description of how CO is transported in space and time in the absence of sources and sinks.

The basic form of a CTM would result if terms corresponding to chemical reactions and sources were added to the right-hand side of equation (3). However, as mentioned in the Introduction, such a model can become quite complex, and more importantly, the final results may contain uncertainties deriving from the dynamical model rather than the true process. For example, the vertical wind $w$, which is known to be poorly observed [see Chapter 3.5 in Holton (2004)], may introduce additional uncertainty in the CO field estimates.

Motivated by this fact, vertical dependence is modeled as a spatial autoregressive term. We make the assumption that the joint distribution of $X_k(L), \ldots, X_k(2), X_k(1)$ can be factorized as

$$[X_k(L), \ldots, X_k(2), X_k(1)] = [X_k(L)|X_k(L-1)] \cdots [X_k(2)|X_k(1)][X_k(1)].$$

Formally, layer $L$ appears to depend only on the layer $L-1$ below it. However, this form is actually symmetric and provides a compact model for the vertical spatial dependence. In fact, this factorization could be reversed, which would imply a forcing from above, instead of from below. This would yield the same joint distribution, which is illustrated with the following example.

Consider a single longitude-latitude coordinate at all layers, at a single time point. Our assumption can then be formalized as an AR(1) model,

$$X(s,l) = fX(s,l-1) + \eta(s,l),$$

where the forcing parameter $f$ is the AR(1) coefficient, and $\eta$ the innovation process. In this model, $X(s,l)$ will have equal correlation with $X(s,l-1)$ and $X(s,l+1)$. This implies that the process simulated under this assumption has the same joint distribution as the following process, which is forced from above: $X(s,l) = fX(s,l+1) + \eta(s,l)$. Furthermore, since the estimate of $X(s,l)$ (for $1 < l < L$) is based on its full conditional distribution, which is proportional to $[X(s,l+1)|X(s,l)][X(s,l)|X(s,l-1)]$, it will depend on the concentrations at the surrounding levels.

For the spatial scales considered here, the main part of the transport is described by the advective terms [the right-hand terms of the top line in equation (3)], and the contribution of the diffusion is expected to be small. Therefore, $\Theta(s)$ is set equal to zero in equation (3), and the remaining small scale variations are approximated with a stochastic innovation process; see below.

2.2.1. *Discretized process model with stochastic terms.* Omitting the diffusion and the vertical contribution in equation (3), the remaining horizontal advection terms are discretized using the Euler step in time and centered differences in space, as described in Haberman (1987). For $l = 2, \ldots, L$ and $k = 1, \ldots, T-1$, this gives

$$
\begin{aligned}
X_{k+1}(s,l) = &\frac{\Delta_k}{2\Delta_x} u_k(s,l)(X_k(s+\Delta_x,l) - X_k(s-\Delta_x,l)) \\
&+ \frac{\Delta_k}{2\Delta_y} v_k(s,l)(X_k(s+\Delta_y,l) - X_k(s-\Delta_y,l)) \\
&+ m(l) X_k(s,l) + f(l) X_k(s,l-1) + \eta_k(s,l),
\end{aligned}
\tag{4}
$$

where $\Delta_x$ is the longitudinal spacing, $\Delta_y$ the latitudinal spacing, and $\Delta_k$ the temporal spacing. The stochastic parameters $m(l)$ and $f(l)$ represent the persistence and forcing parameters, respectively. When equation (3) is discretized, the persistence parameter is modified from being equal to one, to be a stochastic parameter. This makes the model more flexible. Finally, $\eta_k(s,l)$ is assumed to be independent noise distributed as $N(0, \sigma_\eta^2(l))$.

The spatio-temporal neighbors of $X(s, l, t_{k+1})$, $\{X(s+\Delta_x, l, t_k), X(s-\Delta_x, l, t_k), X(s+\Delta_y, l, t_k), X(s-\Delta_y, l, t_k)\}$, have space, time, and wind dependent AR(1)-parameters, that is, $\Delta_k/2\Delta_x u_k(s,l)$ and $\Delta_k/2\Delta_y v_k(s,l)$. The persistence, forcing, and advection terms are believed to contribute to the main part of the transport. Remaining small scale variations are assumed to be modeled by $\eta$.

Discretizing equation (3), using the Euler step in time and centered differences in space, results in an unconditionally unstable solution. That is, letting the process evolve without constraints will result in unbounded state vectors. The choice of using these discretization schemes is to facilitate statistical implementation since it results in a linear system. It should be noted that more accurate discretizations of these equations can be used but will add more complex full conditional distributions. Due to its instability, the model is intractable to use for forecasting, but, when the model is used for interpolation purposes, data will constrain the solution. This distinction has been noted by other researchers in data assimilation; see Wikle et al. (2003).

Finally, the conditional process model for $X_k(l)$, for $l = 2, \ldots, L$ and $k = 1, \ldots, T-1$, is

$$
\begin{aligned}
&X_{k+1}(l) | X_k(l), X_k^B(l), X_{k+1}(l-1), \Theta \\
&\quad \sim MVN(m(l) X_k(l) + f(l) X_{k+1}(l-1) + A_k X_k(l) + A_k^B X_k^B(l), \sigma_\eta^2(l) I),
\end{aligned}
$$

where $X_k^B(l) = \{X_k^B(s_i, l), j = 1, \ldots, N^B\}$ is a boundary process as defined below. The sparse (band-diagonal) matrices $A_k$ and $A_k^B$ contain the AR(1) parameters described above. The initial field, $X_1(1)$, could be assumed to follow a stationary distribution that describes the CO climatology for the given area at that level. Given that, the bottom layer is then simulated [with $f(1) \equiv 0$], then layer two [with $X_{-1}(2) \equiv 0$], and so on.

2.2.2. *Boundary process.* It is clear from equation (4) that values at neighboring locations must be defined. However, note that for a location on the western boundary, for example, there is no western neighbor. Our approach is that the locations at the boundaries are given neighbors that are defined by the so called boundary process, $X^B$. The idea of using stochastic boundary processes in hierarchical dynamical models was originally introduced in Wikle, Berliner and Milliff (2003). The dynamics of the boundary process are chosen to follow a simple random walk,

$$X_{k+1}^B(l) = X_k^B(l) + \eta_k^B(l),$$

where $\eta^B(l) \sim MVN(0, \sigma_B^2(l)I)$ with variance $\sigma_B^2(l)$.

2.3. *Parameter stage.* For each level, the unknown parameters to be estimated are as follows: $m(l), f(l), \sigma_\eta^2(l)$, and, $\sigma_B^2(l)$. For the persistence and forcing parameters, the prior distributions were chosen to be $N(m_0, \sigma_m^2)$ and $N(f_0, \sigma_f^2)$, respectively. These parameters could be designed to be spatially dependent parameters. However, for our application it was decided that our simple choices were found to be sufficient. The prior distributions for the variances are taken to be inverse gamma, $\sigma_\eta^2(l) \sim IG(q_\eta, r_\eta)$, and $\sigma_B^2(l) \sim IG(q_B, r_B)$. These prior distributions are chosen because they are conjugate, which allows us to derive the full conditionals for all parameters. Choices of parameters are discussed below in Section 3.

2.4. *Computational implementation.* The derivations of the full conditional distributions are available as a supplementary document [Malmberg et al. (2008)]. Given these, samples from equation (1) are obtained using the Gibbs sampler. After convergence of the sample paths, posterior means and standard deviations (among other possible statistics) can be obtained. For choice of initial values, and results, see Section 3.

## 3. Application and comparison.

3.1. *Application.* In this section the hybrid physical/statistical model is used to interpolate synthesized GEO satellite observations from a full CTM simulation of the atmospheric CO states. Observations are generated for an air quality surveillance scenario covering the west-coast region of the United States. It is assumed that the future satellite instrument on board the GEO satellite will be capable of CO retrievals at several vertical layers of the atmosphere, but only for cloud-free locations. The goal is to produce the complete CO concentration fields for all vertical layers over the whole region given the satellite observations for cloud-free pixels.

The Community Atmosphere Model (CAM), with simplified CO chemistry [Arellano et al. (2007)], is the CTM used to generate our ground truth. CAM is a state-of-the-art atmospheric general circulation model developed at NCAR [Collins et al. (2006)] for the research community, and is also the atmospheric component of the NCAR climate system model. It is able to simulate the physical and chemical processes of CO, such as emissions, advection, diffusion, deep and shallow convection, boundary layer ventilation, and chemistry. Here, a similar setup for CAM as in Arellano et al. (2007) is used to simulate a realistic spatiotemporal CO process. Being a complete simulation of the atmosphere, CAM also creates clouds and these are used to determine the cloud-free pixels at any given time.

The original horizontal resolution of global CAM data (i.e., output of CAM simulation) is 2.0 degrees latitude $\times$ 2.5 degrees longitude and 26 vertical layers from the surface up to 4 hPa (about 35 km above the surface). In order to simulate a more regional scenario, the CAM data is interpolated to a resolution of 1 degree latitude $\times$ 1 degree longitude and we selected our desired spatial domain from 231.5 to 248.5 degrees longitude and from 33.5 to 48.5 degrees latitude. We also interpolated through the 26 CAM vertical layers and selected 5 vertical layers with 100 hPa thickness centered at 850, 750, 650, 550 and 450 hPa. These levels roughly correspond to levels where current satellite instruments can obtain useful estimates of CO concentrations; see Pan et al. (1998) and Deeter et al. (2004). The CAM simulation of the atmosphere and CO dynamics is based on a 30 minute time step. This compares to a 3 hour interval for the synthetic satellite data.

The initial condition for the simulation of CO in CAM is based on a scenario for a large-scale fire in Southeast Asia for April 2006. The fire is the only source of CO in this study. The winds used for the transport were calculated using CAM, and the same wind fields used in the statistical model. Because CO has a lifetime of about 2 months, such a large fire can add to the local CO concentrations in the Western US, especially above the boundary layer at the lower atmosphere. Emissions of CO from the fire are transported by CAM dynamics, physics, and chemistry, for 24 days, before 4 days of 3 hourly synthesized satellite observations are simulated using equation (2). A location is defined as cloudy if its CAM vertical integrated cloud cover fraction is between 0.5 and 1.

To initialize the Gibbs sampler, $X$ and $X^B$ are simulated from Gaussian distributions with expectations taken as the mean of the observations, and with covariances describing large-scale features that are expected in the complete fields. For every level, the prior means for $m(l)$ and $f(l)$ were chosen to be $m_0 = 1$ and $f_0 = 0$, respectively. The first choice is motivated by the fact that $m(l)$ equals 1 in the deterministic case, as mentioned in Section 2.2. The second choice is motivated by the belief that the main contribution to the transport is across the horizontal dimensions. However, in order to let the observations inform the final estimates, both are chosen to be $\sigma_m^2 = \sigma_f^2 = 10^{-3}$, which compared to the information in the data is a wide variance. The hyperparameters for the variances were set to $(q_\eta, r_\eta) = (2.8, 0.28)$, and $(q_B, r_B) = (2.8, 0.28)$, which corresponds to relatively vague prior knowledge. Given these initializations, 6000 samples from the full conditionals were simulated. This takes about two days. The burn-in time is about 150 samples and the converged chains are thinned before any statistics are calculated.

3.2. *Interpretation of results.* Figure 1 shows samples from prior and posterior distributions for the estimated parameters. It is clear from the shift in the posterior samples that data do have an impact on the estimates. For the bottom level, 850 hPa, there is no forcing from below, and for this level, $m$ explains more of the transport than at higher levels. A possible explanation for the diversity among the posterior samples for $m$ and $f$, at different levels, might be that the wind fields, upon which the physical terms, $A$, and $A^B$, depend, are level dependent. This in turn makes $m$ and $f$ also level dependent. The posterior samples proved to be robust against different choices of priors.

The estimated noise parameters, $\sigma_\eta$ and $\sigma_B$, have a clear trend in the vertical. In this case study CO concentrations are higher in the upper levels, and since the observation noise is proportional to the CO mixing ratio, higher concentrations are observed with larger uncertainty. Hence, less confidence is ascribed to the observations of the upper levels, which in turn inflates the noise parameters as well as their spread.

The top row in Figure 2 shows the synthetic satellite data, $Y$, for time points 10 through 15. Locations marked with a "/" are cloudy and unobserved. Notice how the contiguous patterns of clouds move westward. The second row illustrates the interpolated fields, $\hat{X}$, the posterior mean of $X$, given the observations, at all locations. Posterior means less than zero are marked with a "<" and set to zero for plotting purposes.

Since the estimate for any given level and time point will be informed by data from surrounding levels and time points (where existing), similar to an estimate from a smoother, certain features in the observations may become less apparent in the estimate. For example, the south-east area with high CO concentrations is attenuated. Part of this bias may be due to the nature of the observation noise, which has a standard deviation that is proportional to the true CO concentrations.

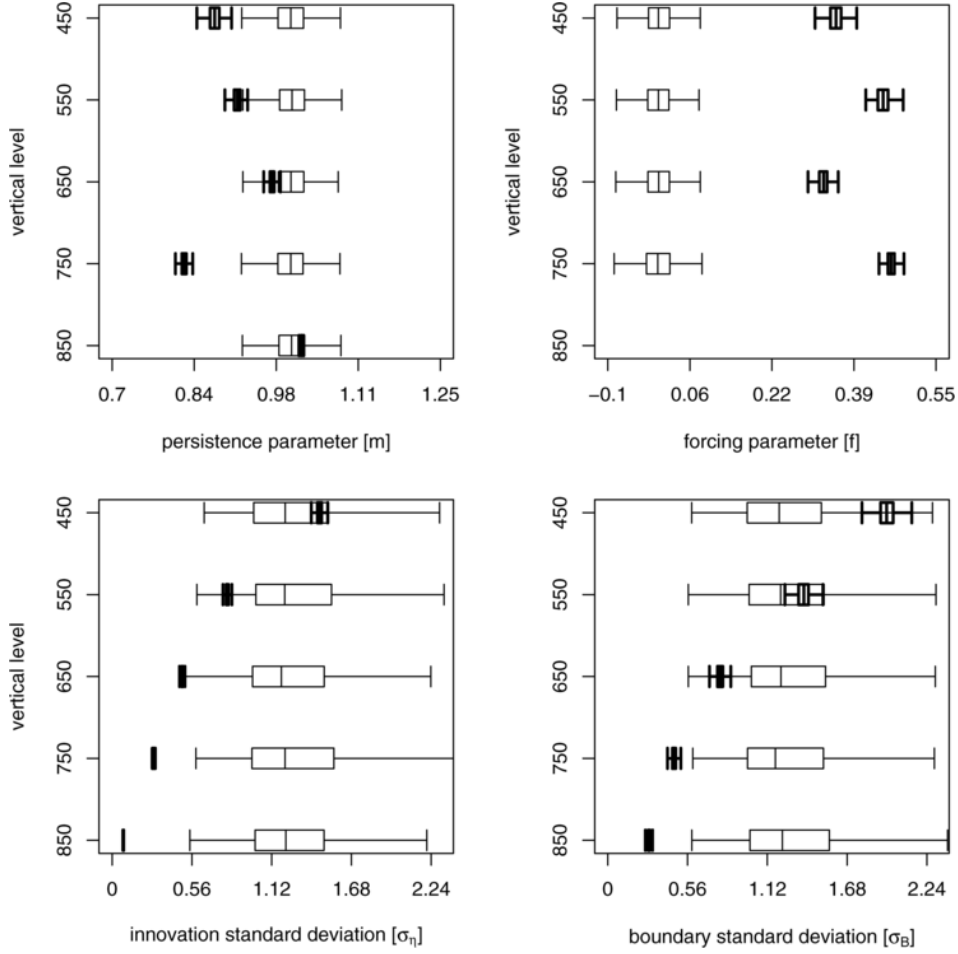

FIG. 1. *Boxplots of prior and posterior samples of estimated parameters. Top-left: Persistence parameter, $m$. Top-right: Forcing parameter, $f$. Bottom-left: Innovation standard deviation, $\sigma_\eta$. Bottom-right: Boundary standard deviation, $\sigma_B$. The gray boxplots represent samples from their prior distributions, and the black boxplots represent samples from their posteriors.*

The third row illustrates the internal estimate of the prediction error, provided by $\hat{\sigma}_X$. Compared to the observed levels, the standard deviations are higher where higher concentrations are observed. Again, this may be due to the nature of the observation noise. In unobserved areas, for example, the north-east area, the uncertainty grows as there are no observations to constrain the numerical error growth in the dynamical model.

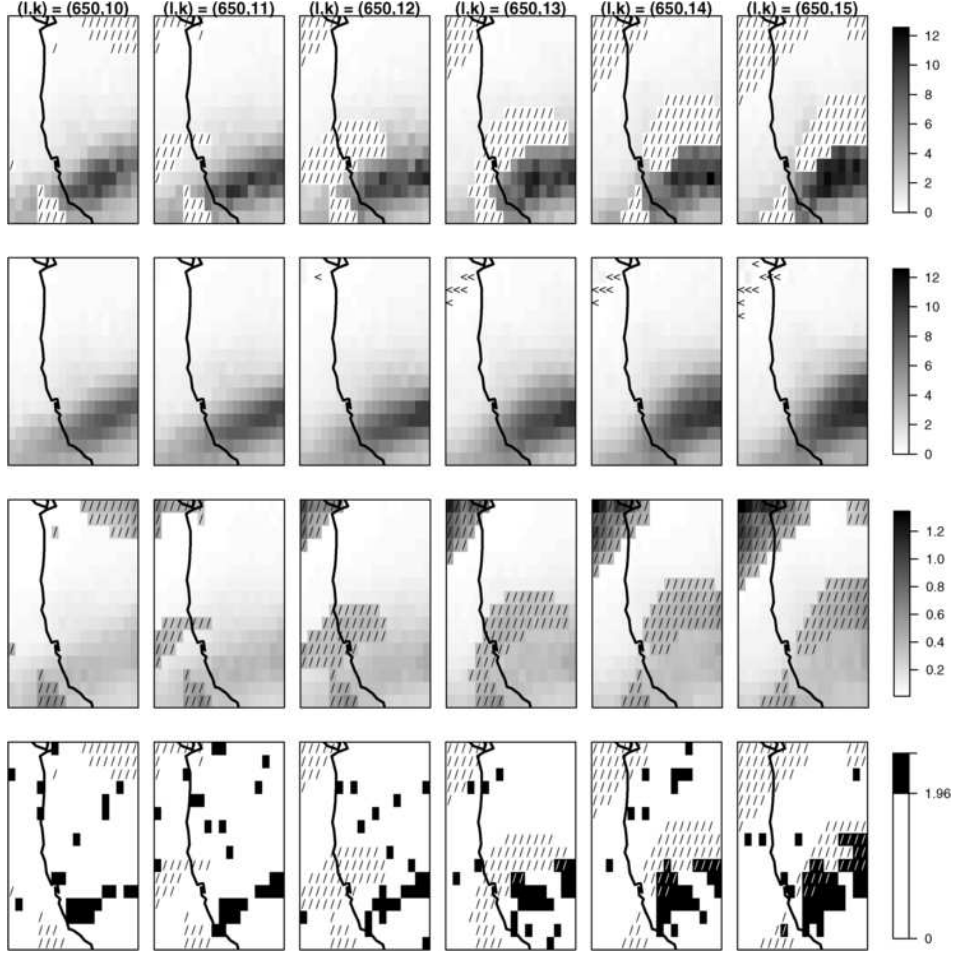

FIG. 2. *Top row: Simulated satellite observations, $Y$, at 650 hPa over the west coast of the USA. Unobserved locations are marked with "/". Second row: Interpolations, $\hat{X}$, where estimates at locations marked with an "<" resulted in a negative estimate. Third row: The internal estimate of the prediction error, provided by $\hat{\sigma}_X$. Bottom row: Standardized residuals, $\{(X - \hat{X})^2/\hat{\sigma}_X^2\}^{1/2}$. Each column corresponds to a time step, and between each column it is three hours. Negative estimates are set to zero, and the color scale reports values of ppb.*

The fourth row is a comparison of the prediction error with $\hat{\sigma}_X^2$. Plotted are the standardized residuals,

$$((X(s,l,k) - \hat{X}(s,l,k))^2/\hat{\sigma}_X^2(s,l,k))^{1/2}.$$

In places with high levels of CO the standard errors tend to underestimate the error. We attribute this to a combination of the smoothing effect and the high observation error. However, overall, the standard errors give reasonable measures of uncertainty even in areas that are masked by clouds and so not directly observed.

As Figure 3 shows, the concentration of CO varies across the vertical. To gauge if the vertical coupling caused too much smoothing, or bias, an additional interpolation was done where $f$ was set to 0. We termed this the uncoupled model. Also, to gauge the spatial prediction skills of these models, which both emphasize the dynamical structure, they were compared to a simple Kriging method; see Cressie (1993). That is, for any level and time point, the process is estimated given the observations at that instance. The Matérn covariance function with smoothness parameter 2.5 is used, and for each interpolation, the variance and range are estimated using maximum likelihood. It should be noted that the spatial smoothing induced by this model is substantially higher than for the dynamical model.

As an objective analysis of these models, the root mean-squared-error ($RMSE$) is calculated,

$$RMSE(l,k) = \left(\frac{1}{N}\sum_{i=1}^{N}(X(s_i,l,k) - \hat{X}(s_i,l,k))^2\right)^{1/2}.$$

The $RMSE$ statistics for level 650 hPa are shown in Figure 4. The performances of the BHMs are worse at the beginning and end of the time period, where less data constrain the interpolations. Furthermore, at the end of the time period, a persistent cloud cover over the north-east area probably contributes to the increased error in the tail. Here the numerical errors in the state vectors are allowed to grow without being constrained by observations. However, where observations constrain the interpolation, the spatial prediction skills are comparable to the Kriging method. Furthermore, the dynamical model makes it possible to compute forecasts.

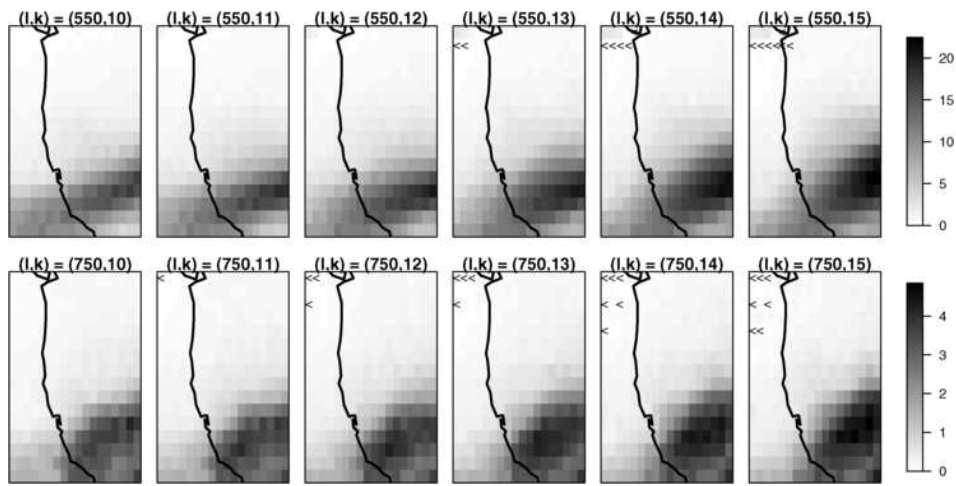

FIG. 3. *Interpolations of the two layers at 550 and 750 hPa. Notice how the concentrations increase with altitude. Color scale reports values of ppb.*

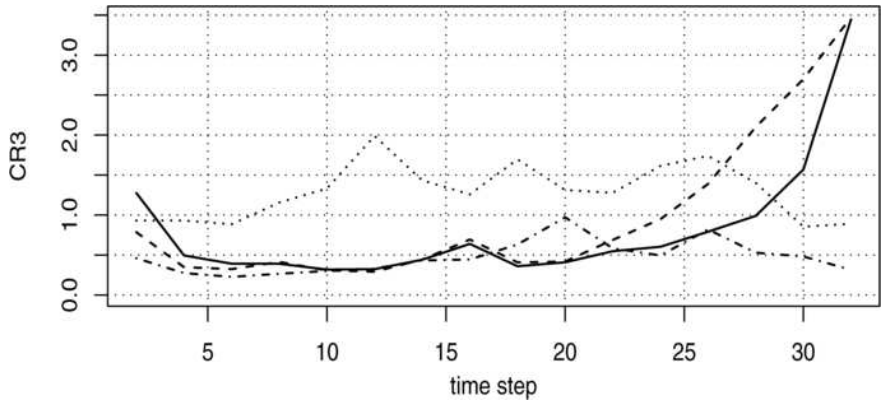

Fig. 4. *The RMSE statistics for level 650 hPa. Solid line: Coupled model; dashed: uncoupled; dot-dashed: Kriging; dotted: CAM/DART.*

In Table 1 the *RMSE* statistics for level 650 hPa are averaged across time. To validate the prediction skills, the statistics were computed for the unobserved (cloudy) locations as well as for all locations. Here, it shows that the coupling is favored. Since the uncoupled model uses less data to compute the interpolations, it will have more spread in its simulations, and a higher *RMSE* than the coupled model. Due to the numerical error growth in the tails, the Kriging method has a lower *RMSE* on the average. A comparison with CAM/DART is presented in the next section.

3.3. *Comparison with CAM/DART.* CAM has most recently been coupled with an ensemble-based data assimilation system, the Data Assimilation Research Testbed (DART), also being developed at NCAR. A description of DART and its evaluation with aircraft measurements is described in Arellano et al. (2007). The present CAM/DART setup has been shown to provide significant improvements in forecast skill of global CO concentrations using joint assimilation of meteorological observations from exist-ing meteorological network and satellite retrievals [Arellano et al. (2007)]. In this study we used the CAM/DART system as our full DA system, assimilating the same synthesized observations as in the statistical model. It should be noted that CAM/DART is a complex software environment and runs reported here require several orders more of computational resources than the BHM. Moreover, measures of uncertainty are not as well developed as in the BHM.

TABLE 1
*RMSE for estimated CO concentrations at 650 hPa using four different methods*

| Method | Unobserved locations | All locations |
|---|---|---|
| Kriging | 0.74 | 0.47 |
| Un-coupled | 1.56 | 0.98 |
| Coupled | 1.28 | 0.81 |
| CAM/DART | 1.10 | 1.29 |

In order to shift CAM away from the configuration used to simulate the ground truth, CAM/DART is initialized with perturbed CO concentrations. This is done by systematically overestimating the emissions of CO from the large fire. We have also used slightly perturbed meteorological conditions for the DA system. As such, the initial CO fields, used in CAM/DART prior to the start of the assimilation experiment, are significantly different from the ground truth, particularly over our spatial domain.

Returning to Table 1, we note that CAM/DART has a relatively higher *RMSE* than the hybrid physical-statistical model for all locations. This indicates that the statistical model has a spread in its prior, such that observations have an impact on the estimate. Conversely, CAM/DART might have a too tight spread in its prior, such that observations are not assimilated properly. An explanation for this might be CAM/DART's spatio-temporal resolution which is aimed for global applications. Typically, a large-scale model does not model small-scale variability, which might be what is missing here. However, CAM has lower *RMSE* for the unobserved locations. This is interpreted as being an artifact of the numerical errors present in the statistical model. Nevertheless, these results show that the statistical model performs reasonably well relative to a full DA system like CAM/DART.

**4. Discussion.** There is strong interest in the future satellite monitoring of atmospheric pollution, such as CO, from the vantage of GEO. This paper is motivated by the fact that these observations will be affected by cloud cover and we present a novel approach to estimating the complete CO distribution based on the available cloud-free data.

The CO fields have dimensions in the horizontal, the vertical, and time. To our knowledge, such a physical/statistical 4-dimensional model has not been considered before. Under the constraints given on our test case, the comparison suggests that the model is comparable to existing methods such as the more complex CAM/DART system. The comparison with CAM/DART is, to our knowledge, the first time a Bayesian spatiotemporal method has been compared directly to a global data assimilation system. We believe it is a significant result that our relatively simpler hybrid physical-statistical model has comparable performance to CAM/DART and the results further provide motivation for the use of BHMs for regional to local applications. We also note that the spatial Kriging method is competitive, but, as we note below, the dynamical BHM can easily be extended to include the kind of spatial dependence used in the Kriging method.

The BHM used in this work was chosen partly for its computational benefits. The additive models in the observation and process levels and the choice of Gaussian innovations has simplified sampling from the posterior because the full conditional distributions have closed forms. Given these choices, the performance of this model is striking and we suspect that it can be improved through several modifications.

Concentrations cannot be negative, but, because of the additive structure of the stochastic components in the BHM, the state variable for CO can be negative. A better approach may be to introduce multiplicative innovations in the process model to preserve positivity. We note that in our current model negative concentrations occur in regions of low concentrations. However, in many cases one is more interested in studying areas with higher concentration and the risk of exceeding some threshold there.

The stochastic component of the process model ($\eta$) has been assumed to be independent in time and space and, thus, the space and time dependence is derived completely from the dynamical form. It is reasonable to consider including some dependence among the components of the $\eta$ process. In particular, Markov random field models would be computationally efficient and provide flexible ranges of spatial dependence. One could also include temporal dependence so that the process model has a moving average term reflecting persistent departures from the dynamics over several time periods. With both of these extensions, the BHM should be able to reproduce the kind of smooth spatial predictions provided by the Kriging approach, and also take advantage of dynamical information. The coupling in the vertical proved to give some improvements over the uncoupled model. However, in this application, the levels are separated by 100 hPa, a separation where the vertical layers can be quite different. It might be that the coupling would prove to be even better if the vertical layers were more dense. Concerning the boundary process, it would be possible to assume a spatial correlation between $X$ and $X^B$, that is, the conditional model for $X^B_{k+1}(l)$ could be extended to include $X_k(l)$.

In addition to modifying the stochastic components, the dynamical model can also be improved. The Eulerian finite differences scheme allowed for simple full conditional distributions, but has the disadvantage of not being stable. Although the stability is constrained based on conditioning the state with observations, at the edges of the time series, where less data is available (in time), the numerical error growth becomes a problem. Higher order and stable difference methods such as a Runge–Kutta scheme would improve the process model. For example, for smooth fields a higher order method will be a more accurate solution to the transport of CO with a changing wind field.

From a scientific point of view, GEO observations contain two problems. The first problem, estimating data at missing locations, is treated here. The second problem, the problem that the observations are correlated in the vertical, has not been addressed. This problem relates to the observation operator $D_k$, which here is a simplified version of a realistic observing system. The BHM facilitates designing $D_k$ such that the vertical correlation is taken into account.

The DART system is designed to accommodate many different types of models and has the capability of approximating an ensemble Kalman smoother [Khare et al. (2008)] as well as a filter. This option is currently not available for CAM but might give better predictions. However, it should be noted that data assimilation systems using a smoother are a rare capability and almost all global systems use filters. Perhaps a more appropriate comparison to a full DA system would have been to use a CTM developed for regional applications, such as WRF-Chem [Grell et al. (2005)], although these models would require information from global models at their boundaries. If a BHM can be formulated within the general ensemble Kalman smoother framework, then DART could be used as the computational engine for approximate sampling of the posterior. This marriage would have substantial advantage, as DART is engineered for large multiprocessor systems, has good control of input and output streams, and would facilitate incorporating more sophisticated physical models at the process level. In this way one could continue exploit advantages of combining both physical models and stochastic elements to improve the estimates of transport.

**Acknowledgments.** The following people have contributed with comments and insights that improved the paper: Reinhard Furrer, Colorado School of Mines; Tim Hoar, National Center for Atmospheric Research; Cari Kaufman, National Center for Atmospheric Research; Finn Lindgren, Lund University; David Moore, King's College London.

We would also like to thank the editor and the anonymous referees for comments that improved the quality of the paper.

## SUPPLEMENTARY MATERIAL

**Full conditional distributions** (DOI: 10.1214/08-AOAS168SUPP; .pdf). The supplement contains notes on the derivations of the conditional distributions appearing in the paper.

A. Malmberg
Ferring Pharmaceuticals A/S
Kay Fiskers Plads 11
DK-2300 Copenhagen S
Denmark
E-mail: anders.malmberg@ferring.com

A. Arellano
Atmospheric Chemistry Division
National Center for Atmospheric Research
Boulder, Colorado 80307
USA
E-mail: arellano@ucar.edu

D. P. Edwards
Program for Atmospheric Composition
  Remote Sensing & Prediction
Atmospheric Chemistry Division
National Center for Atmospheric Research
PO Box 3000
Boulder, Colorado 80307
USA
E-mail: edwards@ucar.edu

N. Flyer
D. Nychka
Institute for Mathematics Applied
  to the Geosciences
National Center for Atmospheric Research
1850 Table Mesa Dr.
Boulder, Colorado 80305
USA
E-mail: flyer@ucar.edu
       nychka@ucar.edu

C. K. Wikle
University of Missouri
146 Middlebush Hall
Columbia, Missouri 65211
USA
E-mail: wiklec@missouri.edu